# High-flux dual-phase percolation membrane for oxygen separation

*Shu Wang[a], Lei Shi[a], Zhiang Xie[a], Yuan He[a], Dong Yan[a], Man-Rong Li[b], Juergen Caro[c], Huixia Luo[a]\**

[a]School of Material Science and Engineering and Key Lab Polymer Composite & Functional Materials, Sun Yat-Sen University, No. 135, Xingang Xi Road, Guangzhou, 510275, P. R. China

[b]School of Chemistry, Sun Yat-Sen University, No. 135, Xingang Xi Road, Guangzhou, 510275, China

[c]Institute of Physical Chemistry and Electrochemistry, Leibniz University of Hannover, Callinstr. 3A, D-30167 Hannover, Germany

*\*Corresponding author/authors complete details (Telephone; E-mail:) (+0086)-2039386124*

*luohx7@mail.sysu.edu.cn*

**Graphical abstract:**

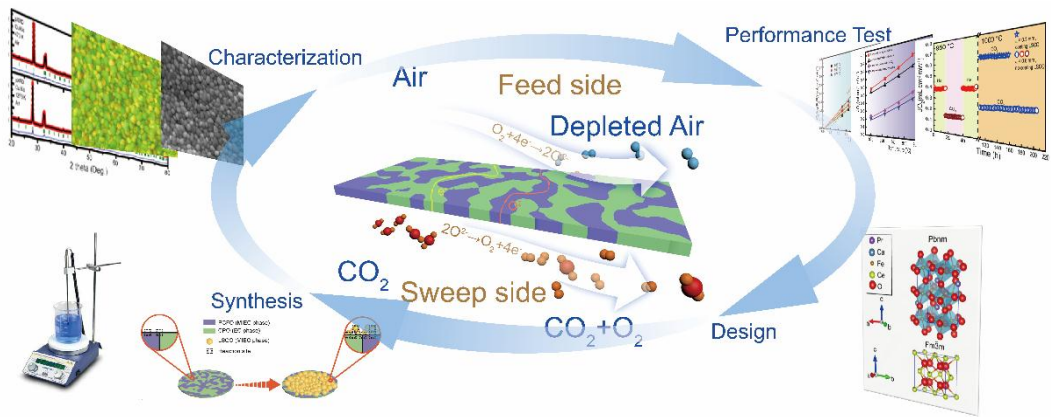

**Abstract:** A series of composites based on $(100-x)$wt.%$Ce_{0.9}Pr_{0.1}O_{2-\delta}$-$x$wt.%$Pr_{0.6}Ca_{0.4}FeO_{3-\delta}$ ($x$ = 25, 40 and 50) doped with the cheap and abundant alkaline earth metal $Ca^{2+}$ at the A-site has been successfully designed and fabricated. The crystal structure, oxygen permeability, phase and $CO_2$ stability were evaluated. The composition of 60wt.%$Ce_{0.9}Pr_{0.1}O_{2-\delta}$-40wt.%$Pr_{0.6}Ca_{0.4}FeO_{3-\delta}$(60CPO-40PCFO) possesses the highest oxygen permeability among three studied composites. At 1000 °C, the oxygen permeation fluxes through the 0.3 mm-thickness 60CPO-40PCFO membranes after porous $La_{0.6}Sr_{0.4}CoO_{3-\delta}$ each to 1.00 mL cm$^{-2}$ min$^{-1}$ and 0.62 mL cm$^{-2}$ min$^{-1}$ under air/He and air/$CO_2$ gradients, respectively. In situ XRD results demonstrated that the 60CPO-40PCFO sample displayed a perfect structural stability in air as well as $CO_2$-containing atmosphere. Thus, low-cost, Co-free and Sr-free 60CPO-40PCFO has high $CO_2$ stability and is economical and environmental friendly since the expensive and volatile element Co was replaced by Fe and Sr was waived since it easily forms carbonates.

# 1. Introduction

Dense ceramic mixed ionic-electronic conducting membranes (MIECMs) are renowned for their rich properties as admirable candidates in energy industries, such as oxygen separation from air, cathode materials for solid fuel oxide cells (SOFCs) or membrane reactors for partial oxidation of light hydrocarbons [1-6]. Especially, the MIECMs integrated oxy-fuel combustion are considered to be a promising cost-effective and efficient method for $CO_2$ capture and sequestration [7, 8]. Membrane operation in oxy-fuel combustion involves exposure to corrosive flue gases such as $CO_2$ and $SO_2$ environments. However, this technology is obstructed since most OTMs suffer from no permeation in the presence of corrosive flue gases such as $CO_2$ and $SO_2$ [9-12].

It is well-known that, according to the phase numbers, the OTMs can be classified into single-phase and composite OTMs. The typical single perovskite phase OTMs (such as $SrFe_{0.8}Co_{0.2}O_{3-\delta}$) have relatively high oxygen permeability, but poor stability under $CO_2$ and reducing atmosphere [13, 14], which limits their application over a wide range. Dual-phase OTMs typically consisted of a fluorite phase ($CeO_2$ types) and a perovskite phase ($ABO_3$ types), which display a better structural and mechanical stability compared with the single-phase OTMs under reducing atmosphere due to the incorporation of fluorite phase. Fluorite phase is usually doped with rare earth elements in the $A$ position (*e.g.* $Ce_{0.9}Nd_{0.1}O_{2-\delta}$) to increase the ionic conductivity [15-17], whereas the perovskite phase is usually doped with rare earth elements in the $A$ site (*e.g.*

$Sm_{0.6}Sr_{0.4}Al_{0.3}Fe_{0.7}O_{3-\delta}$) to enhance stability under $CO_2$ and reducing atmosphere [18-20].

In recent years, considerable efforts have been made to the discovery and synthesis of dual-phase OTMs using various methods, including mixed two phases by hand, ball milling, or one-pot sol-gel method [21-26]. Up to now, abundant composite membrane materials have been reported, such as 60wt.%$Ce_{0.9}Gd_{0.1}O_{2-\delta}$-40wt.%$Ba_{0.5}Sr_{0.5}Fe_{0.2}Co_{0.8}O_{3-\delta}$ [24], and 60wt.%$Ce_{0.9}Gd_{0.1}O_{2-\delta}$-40wt.%$La_{0.6}Sr_{0.4}CoO_{3-\delta}$ [26]. However, according to the Lewis theory of acid-base, most of the aforementioned composite membranes containing strontium or barium alkaline earth elements often form carbonate under $CO_2$ environment since $CO_2$ is acidic gas [27, 28]. Besides, it has been considered that the formation of $CaCO_3$ is much harder than those of $BaCO_3$ and $SrCO_3$ under similar conditions as $Ca^{2+}$ has much smaller ionic radii than those of $Ba^{2+}$ and $Sr^{2+}$ [29]. Recently, $Ca^{2+}$ doped OTMs have been reported to show high $CO_2$ stability due to the lower decomposition temperature of $CaCO_3$ (~ 800 ºC at 1 atm $CO_2$ partial pressure) [29-32]. Further, the *B* site of perovskite phase is usually doped with cobalt to increase the oxygen permeation. However, cobalt is expensive and rare (the average content in the earth's crust is just 0.001%) [33], cobalt and its compounds are also cancer-promoting toxic substance [34].

Therefore, in this work we designed a series of Ca-containing and Co-free dual-phase OTMs (100-*x*)wt.%$Ce_{0.9}Pr_{0.1}O_{2-\delta}$-*x*wt.%$Pr_{0.6}Ca_{0.4}FeO_{3-\delta}$ ((100-*x*)CPO-*x*PCFO) (*x* = 25, 40 and 50; denoted 75CPO-25PCFO, 60CPO-40PCFO and 50CPO-50PCFO, respectively) based on the following aspects: (i) Based on the Ellingham diagram, *e.g.* given by Efimov et al. [29], the formation of $CaCO_3$ is much harder than those of $BaCO_3$ and $SrCO_3$ under similar conditions. (ii) Many researchers have reported that Fe-based OTMs are more stable than Co-based OTMs under reducing atmospheres [35, 36]. In addition, as we all know, abundant iron element exists on the earth, which is much cheaper than Co and suitable for industrialisation. (iii) $Pr_{0.6}Ca_{0.4}FeO_{3-\delta}$ (PCFO) shows fairly high electronic conductivity in the Ca-based perovskite materials, which is beneficial for the oxygen permeability according to the Wagner equation [37]. (iv) $Ce_{0.9}Pr_{0.1}O_{2-\delta}$ (CPO) with good oxygen ionic conductivity and phase stability has been widely used in the fields of SOFCs and oxygen storage [15, 17, 38]. Therefore, based on the previous studies [15, 17, 29, 39], we propose to replace Co by Fe and Sr by Ca. Besides, the oxygen permeability of a fluorite-perovskite OTM usually depends on the transport rate of oxygen ions due to oxygen ions transport much slower than electrons [40, 41]. Under the premise of two phases form respectively interpenetrating continuous network, it is reasonable to maximize the content of fluorite phase when designing a fluorite-perovskite OTM.

Here we develop a series of the (100-$x$)CPO-$x$PCFO ($x$ = 25, 40 and 50) OTMs, which were synthesized by a Pechini one-pot method. The effects of different CPO/PCFO ratios on the crystal structure, morphology, oxygen permeability and structural stability as well as $CO_2$ stability are investigated. We will focus on the phase stability of PCFO and CPO in $CO_2$ and low oxygen partial pressure atmospheres.

## 2. Experimental

2.1. Powder and membrane preparation

The (100-$x$)CPO-$x$PCFO ($x$ = 25, 40 and 50) composites powders were prepared by a Pechini one-pot method with only citric acid as the chelating agent. As presented in **Fig. S1**, the stoichiometric amount of starting materials were dissolved in a solution with the ratio of citric acid to metal cations of 2 : 1. At the meantime, a small amount of ethylene glycol as the surfactant was added to the solution with stirring in the beaker. A gel was formed after the solution was gradually evaporated and heated at 110 °C. A precursor powders can be obtained after heating in an oven at 150 °C. The precursor powders were then calcined in air at 600 °C for 8 h to remove the organic compounds and obtain the primary phase. The dual-phase powders can be obtained after heating in air at 950 °C for 10 h. The obtained powders were grounded and pressed into a disk in a die under pressure of ~ 12.5 MPa. The black disks were sintered at 1400 °C for 5 h in air with a heating/cooling rate of 1.5 °C to gain dense membranes. The membranes were carefully polished to be thickness of 0.80 mm, 0.60 mm, 0.50 mm, 0.40 mm and 0.30 mm by using 1200 grit-sand papers, respectively. Finally, the dual-phase membranes were prepared for test after washing with ethanol. The slurry with composition of 60wt.% La$_{0.6}$Sr$_{0.4}$CoO$_{3-\delta}$ (LSCO) and 40wt.% terpineol was deposited on air side of the 0.30 mm-thickness membrane and heated in air for 2 h, and then a porous LSCO layer can be formed, which is taken to enhance the oxygen surface exchange rate on the air side.

## 2.2. Membrane characterizations

The phase structures of powder samples, fresh and spent membranes with composition of (100-$x$)CPO-$x$PCFO ($x$ = 25, 40 and 50) were investigated using room temperature X-ray diffraction (XRD, D-MAX 2200 VPC, Rigaku with Cu Kα). For *in situ* XRD, the 60CPO-40PCFO powders were also conducted between room temperature and 1000 °C in air or 100 mL min$^{-1}$ 60vol. % $CO_2$/40 vol.% $N_2$ with a heating rate of 10 °C min$^{-1}$ and relaxation time of 30 min at each temperature before measurements. To determine the unit cell parameters, Rietveld fitting was performed on the powder diffraction data through the use of the FULLPROF diffraction suite [42]. The crystal structures were generated using the Vesta program [43]. The morphologies of the as-sintered membranes and the membranes through heat treatment were studied by scanning electron microscopy (SEM, Quanta 400F, Oxford) operated at 20 kV. The grains composition of the fresh 75CPO-25PCFO membrane was checked by energy dispersive X-ray spectroscopy (EDXS). Transmission electron microscopy (TEM), high-resolution TEM (HRTEM) and selected area electron diffraction (SAED) were used to investigate the crystal structure of the dual-phase material as powder calcined at 950 °C for 10 h in air. All these analyses (HRTEM, SAED) were performed on JEOL 2010F operated at 200 kV.

## 2.3. oxygen permeation measurements

The oxygen permeability and $CO_2$ stability were studied using a homemade corundum reactor as reported elsewhere [38, 39]. The membrane was sealed in an alumina tube with a ceramic sealant (Huitian, Hubei, China). The effective area of membrane in tube is around 0.709 cm$^2$. Synthetic Air (79vol.% N$_2$ and 21vol.% O$_2$) as a feeding gas and He (99.999vol.% He)+Ne (99.999vol.% Ne)/CO$_2$ (99.999vol.% CO$_2$)+Ne (99.999vol.% Ne) as the sweeping gas separately flowed into each side of the membrane. All the gases flow was controlled by the mass flow meter (Sevenstar, Beijing, China). Also, the oxygen permeation process always exits a small amount of oxygen leakage, where this part shall not exceed 10% of the total amount. The leakage of oxygen also should be reduced when calculating the oxygen permeation fluxes. The effluent content was detected by gas chromatography (GC, zhonghuida-A60, Dalian, China), the oxygen permeation rate was calculated according to *eq.* 1: [44-46]

$$J_{O_2}(\text{mL cm}^{-2}\text{ min}^{-1}) = \frac{C_{O_2}}{1-C_{O_2}} \times \frac{F}{S} \qquad \textit{eq. 1}$$

where $J_{O_2}$ represents the oxygen permeation rate, $C_{O_2}$ is the effective oxygen content subtracted by leaking oxygen, *F* is the total flow of effluents and *S* is the effective area sealed in alumina tube.

## 3. Result and Discussion

3.1. Phase structure characterization

The XRD patterns of (100-*x*)CPO-*x*PCFO (*x* = 25, 40 and 50) samples heated at 950 °C for 10 h in air (**Fig. 1**), verify all (100-*x*)CPO-*x*PCFO (*x* = 25, 40 and 50) powders are composed of only the two phases CPO (space group *No.*255: *Fm$\bar{3}$m*) and PCFO (space group *No.*62: *Pbnm*). Top inset of **Fig. 1b** shows the (200) peaks of CPO phase (right peak) and the overlap peaks (left peak) composed of (020) (112) (200) peaks of PCFO phase which are owing to the pseudocubic structure ($a_o \times b_o \times c_o \approx \sqrt{2}a_c \times \sqrt{2}a_c \times a_c$; $a_o$, $b_o$, $c_o$) are the orthogonal phase cell parameters; $a_c$ is the cubic phase cell parameter) [47]. From the intensity change for relative ratios of the left peak and the right peak, we can ensure the relative content change of CPO and PCFO phases. In addition, for comparison, the single-phase CPO and PCFO were also prepared under the same condition. The unit cell parameters of the pure CPO, PCFO and (100-$x$)CPO-$x$PCFO ($x$ = 25, 40 and 50) dual-phase ceramic powders are listed in Table 1. It is clearly seen that the lattice constants for (100-$x$)CPO-$x$PCFO ($x$ = 25, 40 and 50) composite powders are very close to those for the pure CPO (a = b = c = 5.4048(1)Å) and the pure PCFO (a = 5.4456(9)Å, b = 5.4802(9)Å, c = 7.717(1)Å). The slight differences in b-parameter of PCFO and a-parameter of CPO could be ascribed to a migration of Pr from one phase to the other occurring at high temperature. As presented in **Table S1-3**, the ratios of Pr/Ca in three composition membranes (0.642(7):0.358(7), 0.674(7):0.326(7), 0.671(11):0.329(11), respectively) are slightly higher than the setting values. This behavior is also reflected in the EDXS results. For example, as shown in **Fig S2**, the ratio of Pr/Ca of 60CPO-40PCFO membrane is 0.628:0.372, which is also slightly bigger than the setting value. **Fig. S3** presents the XRD patterns of (100-$x$)CPO-$x$PCFO ($x$ = 25, 40 and 50) composites sintered at 1400 °C for 5 h. All composites after sintering consist of only the two phases CPO ad PCFO. Fang *et al*. have reported that Ca-contained composite membrane can easily form secondary phase (such as brownmillerite) by glycine-nitrate combustion synthesis [28, 48]. However, no additional phases, such as $SrPrFeO_4$ and $PrFeO_3$, were detected within the resolution limit of XRD, in our dual-phase membranes obtained by the Pechini one-pot method in this work. Obviously, the dual-phase (100-$x$)CPO-$x$PCFO membranes can be successfully synthesized via this method.

In addition to XRD, HRTEM and SAED of 60CPO-40PCFO powder prepared at 950 ℃ for 10 h were also conducted. **Fig. 2a, b** show the HRTEM images of CPO and PCFO in the composite 60CPO-40PCFO powder, the characteristic (111) CPO and (110), ($\bar{1}$10) PCFO in the 60CPO-40PCFO powder. Fast Fourier transform images (FFT) are shown in the upper right corner of the image. **Fig. 2c, d** show the [$1\bar{1}\bar{1}$] zone axis pattern (ZAP) of CPO and the [001] ZAP of PCFO, respectively. It indicates that CPO phase has a cubic structure; the PCFO has orthorhombic structure, which is consistent with the XRD data previously mentioned.

3.2. Surface morphology characterization

In order to investigate the suitable sintering temperatures of (100-$x$)CPO-$x$PCFO ($x$ = 25, 40 and 50) composite membranes, several sintering temperatures have been performed. We focused on the 60CPO-40PCFO sample. **Fig. S4** shows SEM pictures of the selected 60CPO-40PCFO sample fabricated at different sintering temperatures for 5 h with a heating/cooling rate of 1.5 °C min$^{-1}$. It can be seen that the sample after treating at 1300 °C for 5 h was not dense enough for the later oxygen permeability measurements. However, the grain appeared irregular shape and the 60CPO-40PCFO composite was molten if the sintering temperature was held at 1500 °C for 5 h. This heating/cooling rate also plays an important role in how to get a dense 60CPO-40PCFO membrane. If the heating and cooling rates are too high, the membranes will crack. Thus, the optimal sintering condition was found to be 1400 °C for 5 h with a heating and cooling rate of 1.5 °C min$^{-1}$.

In the following, we study the microstructures of all sintered 50CPO-50PCFO, 60CPO-40PCFO and 75CPO-25PCFO samples at the optimal sintering condition. **Fig. 3** depicts the secondary electron (SE) and backscattered electron (BSE) images of three 50CPO-50PCFO, 60CPO-40PCFO and 75CPO-25PCFO samples. It is seen that all membranes are stacked closely and free of cracks. All composites compose of only CPO and PCFO grains, and the CPO grains are similar to the PCFO grains about 1 μm except for 75CPO-25PCFO. It is seen that there are some grey grains with grain size much smaller than 0.1 μm in the 75CPO-25PCFO membrane. For further study, we performed EDXS in the dark grains and the grey grains in the 75CPO-25PCFO membrane, as shown in **Fig S5 b and c**, it can be found that the dark grains are the component ($Pr_{0.50}Ca_{0.37}Ce_{0.19}Fe_{0.95}O_3$) and grey grains are Ce-rich component ($Pr_{0.47}Ca_{0.2}Ce_{0.68}Fe_{0.64}O_3$). A similar phenomenon recently has been reported by Cheng et al. in the ceramic $H_2$-permeable ceramic membrane composition $BaCe_{0.5}Fe_{0.5}O_{3-\delta}$, which auto-decomposes on heating at 1370 °C for 10 h into the cerium-rich oxide $BaCe_{0.85}Fe_{0.15}O_{3-\delta}$ (BCF8515) and the iron-rich oxide $BaCe_{0.15}Fe_{0.85}O_{3-\delta}$ (BCF1585) two perovskite oxides [49], which was attributed to the large differences between the radii of $Ce^{3+}$ and $Ce^{4+}$ (1.15 and 1.01) and $Fe^{2+}$, $Fe^{3+}$, and $Fe^{4+}$ (0.92, 0.785, and 0.725) [50, 51]. The auto separation of BCF8515 is of great benefit for the $H_2$ permeability.

However, in our 75CPO-25PCFO case, it seems like that the Ce-rich phase would be harmful to the oxygen permeability, which will be described more in detail in the following "oxygen permeation performance" part.

3.3. Oxygen permeation performance

The oxygen permeation fluxes through (100-$x$)CPO-$x$PCFO ($x$ = 25, 40 and 50) composites were performed in an air/He gradient at temperatures between 900 °C and 1000 °C. As shown in **Fig. 4**, increases in oxygen permeation rates with the rising temperature are in the following order: 75CPO-25PCFO < 50CPO-50PCFO < 60CPO-40PCFO. Obviously, the composition of 60CPO-40PCFO possesses the highest oxygen permeability; while the composition of 75CPO-25PCFO has the lowest oxygen permeability during these three compositions, which may be ascribed to the formation of the cerium-rich phase in 75CPO-25PCFO membrane, leading to the block of the electronic conducting path. Further study needs to confirm this behavior.

To study the influence of thickness ($L$) on our (100-$x$)CPO-$x$PCFO ($x$ = 25, 40 and 50) composites, we focused on the optimal 60CPO-40PCFO sample. **Fig. 5a** depicts the influence of the different thickness ($L$) on oxygen permeability of 60CPO-40PCFO membranes in dependence of temperature. All the oxygen permeation fluxes through the 60CPO-40PCFO composites enhance with the enhancing temperatures. *E.g.* the oxygen permeation rate for the 0.30 mm-thickness membrane enhances from 0.53 mL cm$^{-2}$ min$^{-1}$ to 0.87 mL cm$^{-2}$ min$^{-1}$ when the temperature enhances from 900 to 1000 ºC. **Fig. 5b** displays the oxygen permeation fluxes through the 60CPO-40PCFO membranes with different thickness as Arrhenius plot in air/He gradient. Further, the apparent activation energies of 60CPO-40PCFO with the thickness of 0.80 - 0.30 mm in the temperature range of 900 - 1000 °C were calculated to be 74.32, 70.55, 69.89 and 62.92 kJ/mol.

Generally speaking, bulk diffusion and surface exchange constitute the two main processes during oxygen transport through MIECMs. To further understand the rate-determining steps of our 60CPO-40PCFO, we plotted $J_{O_2}/\log(P_h/P_l)$ versus $1/L$ at 925, 950 and 975 °C (**Fig. 6**, data extracted from **Fig. 5a**). It can be seen that when the thickness is greater than 0.30 mm, $J_{O_2}/\log(P_h/P_l)$ is a linear function of $1/L$. This can be described in the *eq.* 2 [52],

$$J_{O_2}(\text{mL cm}^{-2}\text{ min}^{-1}) = \frac{C_V D_V}{4L}\ln\frac{P_h}{P_l} \qquad \textbf{\textit{eq. 2}}$$

where $J_{O_2}$ is the oxygen permeation rate, $C_V$ and $D_V$ are the concentration of oxygen vacancies and the diffusion coefficient of oxygen vacancies. $P_h$, $P_l$, $L$ denote the high oxygen partial pressure on the feed side, the low oxygen partial pressure on the sweep side, and membrane thickness, respectively. It can be signified that if the bulk diffusion is the rate-limiting process, the oxygen permeation rate will have an inversely proportional relationship with $L$ under a constant oxygen partial pressure gradient. If not, the surface exchange will be the rate-limiting process if the membrane thickness is thin enough. From *eq.* 2, we can expect that a plot of $J_{O_2}/\log(P_h/P_l)$ against $1/L$ gives a straight line. However, the oxygen permeation rate starts to shift from the straight line when the thickness reduces to be 0.3 mm, which suggest that the bulk diffusion is the rate-determining step when the thickness is larger than 0.3 mm.

As is presented in **Fig. 7a**, the surface exchange often occurs in triple phase boundary (TPB) sites and the number of amount of TPB sites can be increased by modifying the membrane surface with coating a LSCO porous layer [53, 54]. Therefore, when the surface exchange is the rate-determining step, the oxygen permeation rate of the OTMs can be enhanced by modifying their surfaces. **Fig. 7b** shows the oxygen permeation fluxes through the 0.30 mm-thickness 60CPO-40PCFO composites as a function of temperatures with and without LSCO coating. All the oxygen permeation fluxes through the LSCO-coating membranes with 0.30 mm thickness were enhanced by coating with porous LSCO layer, no matter using He or $CO_2$ as the sweep gas. For example, the oxygen permeation rate enhances from 0.87 mL $cm^{-2}$ $min^{-1}$ to 1.00 mL $cm^{-2}$ $min^{-1}$ at 1000 °C with He as sweep gas.

3.4. Stability study

In addition, the long-term oxygen permeability of the 0.60- or 0.30 mm-thickness 60CPO-40PCFO membranes using $CO_2$ as sweep gas were further conducted to check the $CO_2$ stability. As presented in **Fig. 8,** a stable oxygen permeation flux of 0.38 mL $cm^{-2}$ $min^{-1}$ for the 0.60 mm-thickness 60CPO-40PCFO membrane was obtained under air/He atmosphere. When it switched to $CO_2$ on the sweep side, it immediately decreased to a slightly low value of 0.15 mL $cm^{-2}$ $min^{-1}$ because of the $CO_2$ adsorption effect on the oxygen surface-exchange. Similar behaviors have been found in other $CO_2$-stable composites, *e.g.* the oxygen permeation rate of 60wt.%$Ce_{0.9}Nd_{0.1}O_{2-\delta}$-40wt.%$Nd_{0.6}Sr_{0.4}CoO_{3-\delta}$ membrane with 0.60 mm thickness almost declined from 0.65 mL $cm^{-2}$ $min^{-1}$ to 0.55 mL $cm^{-2}$ $min^{-1}$ when tuning pure He to pure $CO_2$ [55]. Zhu *et al* also found the obvious decline on the 75wt.%$Ce_{0.8}Sm_{0.2}O_{0.19}$- 25wt.%$Sm_{0.8}Ca_{0.2}Mn_{0.5}Co_{0.5}O_{3-\delta}$ membrane, it attributes to stronger adsorption of $CO_2$ on the membrane surface as doping of calcium [56]. In addition, the oxygen permeation rate resumed back to 0.38 mL $cm^{-2}$ $min^{-1}$, if the sweep gas was changed back from $CO_2$ to He. Besides, it can be seen that the oxygen permeation fluxes of the 60CPO-40PCFO membranes with 0.60 mm thickness without coating and 0.30 mm thickness with coating maintained around 0.25 mL $cm^{-2}$ $min^{-1}$ over the 100 h and 0.62 mL $cm^{-2}$ $min^{-1}$ over the 50 h, respectively, and no decrease was observed during the long-term test. This behavior differs from the previous reports on the single-phase membranes. It is well known that many single perovskite OTMs are not stable under $CO_2$-containing atmospheres, resulting in sharp drop of oxygen permeation rate and even break the membranes [57-60]. For instance, no oxygen permeation for $Ba_{0.5}Sr_{0.5}Fe_{0.8}Co_{0.2}O_{3-\delta}$ membrane has been observed when using pure $CO_2$ as sweep gas at 875 °C [57]. More recently, Jian Song et al. has reported that there is no permeation through the single perovskite $BaCo_{0.85}Bi_{0.05}Zr_{0.1}O_{3-\delta}$ hollow fiber samples when even 10 vol.% $CO_2$ in the feed or sweep sides due to the formation of $BaCO_3$ carbonate on the surface [58]. In order to further confirm the operating stability of the 60CPO-40PCFO membrane, XRD was used to check the crystal structures of both sides of the spent 60CPO-40PCFO membrane with 0.6 mm after conducted at over 150 h under Air/$CO_2$ atmospheres at 1000 °C. As shown in **Fig. S6**, the crystal structures of the 60CPO-40PCFO membranes before and after the long-term oxygen permeation measurements with pure $CO_2$ as sweep gas kept unchanged, further revealing the high $CO_2$ stability of the 60CPO-40PCFO composite.

Next, we also consider the $CO_2$ stability of 60CPO-40PCFO at low temperature. Thus, the 60CPO-40PCFO powder is also treated under pure $CO_2$ at 800 °C for 48 h.

For comparison, the Co-doping 60wt.%$Ce_{0.9}Pr_{0.1}O_{2-\delta}$-40wt.%$Pr_{0.6}Ca_{0.4}Fe_{0.8}Co_{0.2}O_{3-\delta}$ (60CPO-40PCFCO) sample has also been treated in the same condition. **Fig. 9** presents the XRD patterns of 60CPO-40PCFO and 60CPO-40PCFCO powders before and after exposing in pure $CO_2$ atmosphere. Insets of **Fig. 9** were the enlarged graph of CPO (100) peak around. Obviously, calcium carbonate phase marked with "*" was observed on the 60CPO-40PCFCO sample after exposing to pure $CO_2$ at 800 ℃ for 48 h, but no carbonate peak was detected for our 60CPO-40PCFO sample within the limit of XRD, which further confirmed the 60CPO-40PCFO composite is $CO_2$ stable at low temperature.

To further characterize the high-temperature structure stability in different atmospheres, 60CPO-40PCFO powder prepared at 950 °C for 10 h was studied by *in situ* XRD. There was no structural phase transition observed in the whole *in situ* XRD test in air (**Fig. 10a**), which suggested the 60CPO-40PCFO composite has high phase stability in air. In order to determine its stability with the existence of $CO_2$, further *in situ* XRD measurement was performed. **Fig. 10d** shows the *in situ* XRD pattern of 60CPO-40PCFO sample in 60vol.% $CO_2$/40vol.% $N_2$ atmosphere upon heating between 25 and 1000 °C. It can be seen that the phases maintain unchanged during heating process, and no carbonate peaks arise, demonstrating that the 60CPO-40PCFO membrane is also stable under a $CO_2$ environment. Meanwhile, the unit-cell parameters for CPO and PCFO as a function of temperature are shown in the **Fig. 10b, c, e, f**. It can be seen that the samples are exposed no matter in air or 60vol.% $CO_2$/40vol.% $N_2$ atmosphere, the unit-cell parameters for PCFO could be divided into two segments: the low (25 - 600 °C) and high temperature (800 - 1000 °C) parts. The slope of the high temperature line is more inclined than that of the low temperature line, which could be ascribed to the formation of more oxygen vacancies apart from lattice expansion in the elevated temperature [45]. As for the two segments of fluorite phase CPO in **Fig. 10c, f**, it could be ascribed to the transformation multivalent Pr (+3/+4) in the fluorite phase, and some similar phenomenon has been reported by Balaguer [16]. We further examine the phase stability of the 60CPO-40PCFO in low oxygen partial pressure by treating the 60CPO-40PCFO powders with pure Ar at various temperatures for 48 h and different times at 950 °C, as shown in **Fig. S7** a and b, respectively. From the identical XRD patterns in **Fig. S7**, we can conclude that the 60CPO-40PCFO shows excellent stability of 60CPO-40PCFO in Ar as well.

Finally, **table 2** compares the oxygen permeability of several types of cobalt-free oxygen permeable composite membranes under air/He or air/$CO_2$ gradient [9, 39, 61, 62]. It can be seen that our membrane (thickness: 0.6 mm) shows comparable or even higher oxygen permeability than those of some composites at 1000 °C, such as 60wt.%$Ce_{0.9}Gd_{0.1}O_{2-\delta}$-40wt.%$Fe_2O_3$(thickness: 0.5 mm), 40vol.%$Ce_{0.9}Tb_{0.1}O_{2-\delta}$-60vol.%$NiFe_2O_4$(thickness: 0.68 mm). When we reduce the thickness of the membrane and conduct surface modification on the membrane surface, oxygen permeability is further improved. The oxygen permeation flux of our membrane (thickness: 0.3 mm; coating: $La_{0.6}Sr_{0.4}CoO_{3-\delta}$; operating temperature: 1000 °C) is 0.62 mL $cm^{-2}$ $min^{-1}$ at air/$CO_2$ gradient, which is nearly 2.5 times greater than that of 60CPO-40PSFO membrane (thickness: 0.6 mm; operating temperature: 950 °C).

## 4. Conclusions

In this work, we have designed a series of novel Ca-containing $CO_2$-stable membranes based on the (100-$x$)CPO-$x$PCFO, which can be prepared successfully by a Pechini one-pot method. The replacement of Co by Fe makes our dual-phase membrane economic and ecologic. The (100-$x$)CPO-$x$PCFO membranes sintered at 1400 ºC for 5 h in air are composed of only CPO and PCFO two phases except for 75CPO-25PCFO. It was found that 60CPO-40PCFO possess the optimal oxygen permeability among three compositions and yields a high oxygen permeation rate of 1.00 mL cm$^{-2}$ min$^{-1}$ for the 0.30 mm-thickness 60CPO-40PCFO membrane after LSCO coating at 1000 °C in air/He gradient. In addition, 60CPO-40PCFO membranes can be operated stably for over 100 h under pure $CO_2$, which suggests that 60CPO-40PCFO membranes have good $CO_2$ stability and are promising for using them in oxy-fuel route for $CO_2$ capture.

**Acknowledgment**

H.X. Luo acknowledges the financial support by "Hundred Talents Program" of the Sun Yat-Sen University and National Natural Science Foundation of China (21701197). M.R. Li is supported by the "One Thousand Youth Talents" Program of China.

# Supporting information

**High-flux dual-phase percolation membrane for oxygen separation**

*Shu Wang[a], Lei Shi[a], Zhiang Xie[a], Yuan He[a], Dong Yan[a], Man-Rong Li[b], Juergen Caro[c], Huixia Luo[a]\**

[a] School of Material Science and Engineering and Key Lab Polymer Composite & Functional Materials,, Sun Yat-Sen University, No. 135, Xingang Xi Road, Guangzhou, 510275, P. R. China

[b] School of Chemistry, Sun Yat-Sen University, No. 135, Xingang Xi Road, Guangzhou, 510275, China

[c] Institute of Physical Chemistry and Electrochemistry, Leibniz University of Hannover, Callinstr. 3A, D-30167 Hannover, Germany

Table S1. Refined PCFO structural parameters of 50CPO-50PCFO dual-phase powder

$R_{wp} = 5.68\%, R_p = 4.54\%, \chi2 = 1.18$

| Atom | x | y | z | Occupancy |
|---|---|---|---|---|
| O1 | 0.214(5) | 0.029(3) | 0.279(5) | 1 |
| Ca1 | 0.4643(5) | 0.25 | 0.001(2) | 0.358(7) |
| Pr1 | 0.4643(5) | 0.25 | 0.001(2) | 0.642(7) |
| O2 | 0.533(5) | 0.25 | 0.629(6) | 1 |
| Fe1 | 0 | 0 | 0 | 1 |

Table S2. Refined PCFO structural parameters of 60CPO-40PCFO dual-phase powder

$R_{wp} = 5.21\%, R_p = 4.12\%, \chi2 = 1.09$

| Atom | x | y | z | Occupancy |
|---|---|---|---|---|
| O1 | 0.216(4) | 0.035(3) | 0.278(5) | 1 |
| Ca1 | 0.4625(4) | 0.25 | 0.002(2) | 0.326(7) |
| Pr1 | 0.4625(4) | 0.25 | 0.002(2) | 0.674(7) |
| O2 | 0.514(4) | 0.25 | 0.613(5) | 1 |
| Fe1 | 0 | 0 | 0 | 1 |

Table S3. Refined PCFO structural parameters of 75CPO-25PCFO dual-phase powder

$R_{wp} = 6.44\%, R_p = 5.08\%, \chi2 = 1.3499$

| Atom | x | y | z | Occupancy |
|---|---|---|---|---|
| O1 | 0.207(5) | 0.036(5) | 0.283(7) | 1 |
| Ca1 | 0.4616(7) | 0.25 | 0.001(3) | 0.329(11) |
| Pr1 | 0.4616(7) | 0.25 | 0.001(3) | 0.671(11) |
| O2 | 0.499(6) | 0.25 | 0.608(8) | 1 |
| Fe1 | 0 | 0 | 0 | 1 |

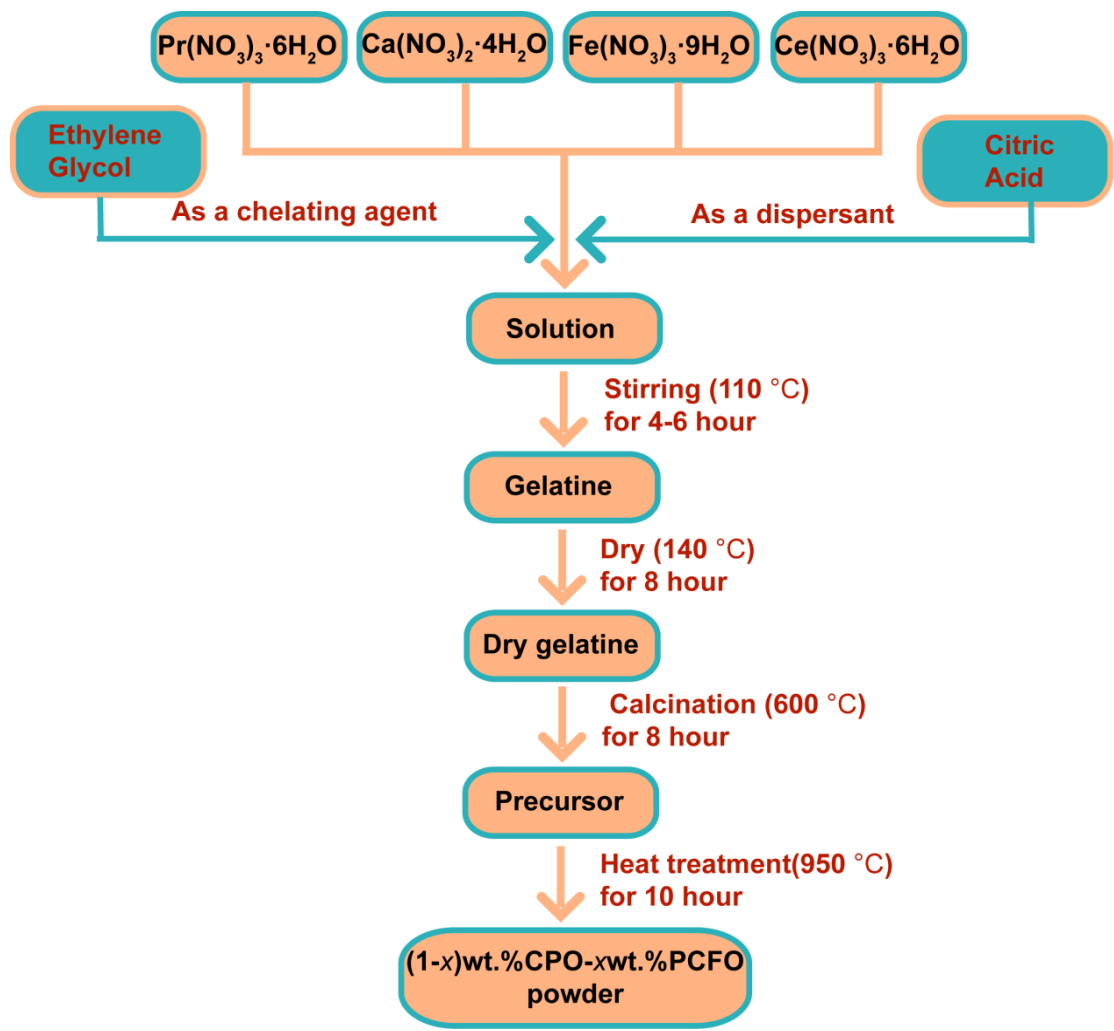

**Figure S1.** Flowchart for synthesis of (100-*x*)CPO-*x*PCFO (*x* = 25, 40 and 50) composite powders via a Pechini one-pot method.

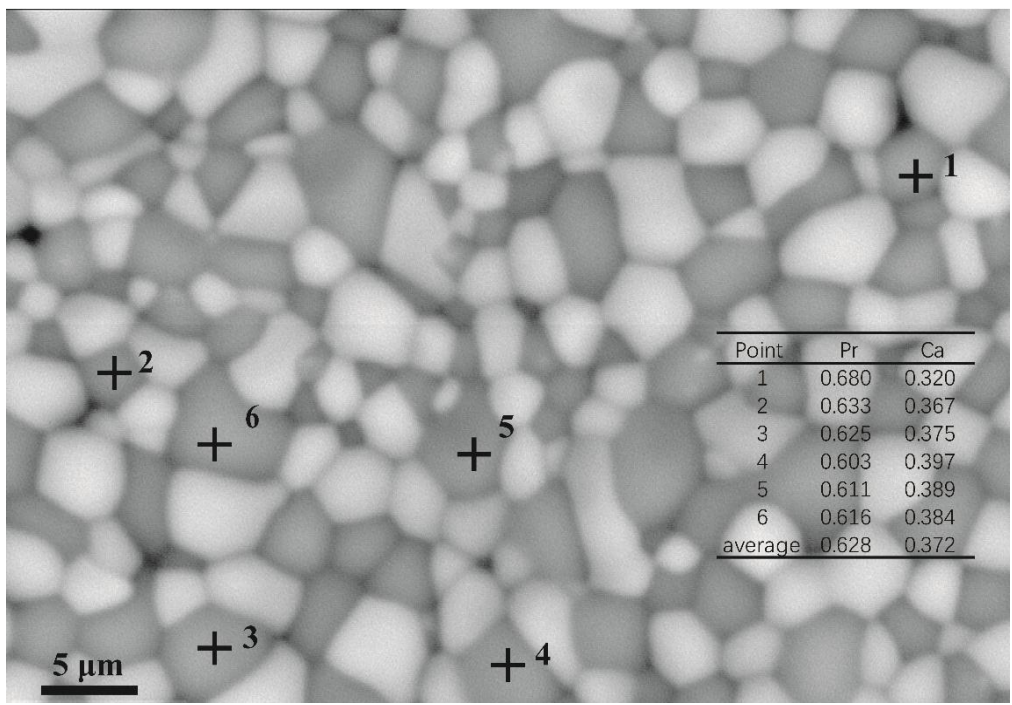

**Fig S2.** EDXS image of 60CPO-40PCFO membrane after sintering at 1400 °C in air for 5 h.

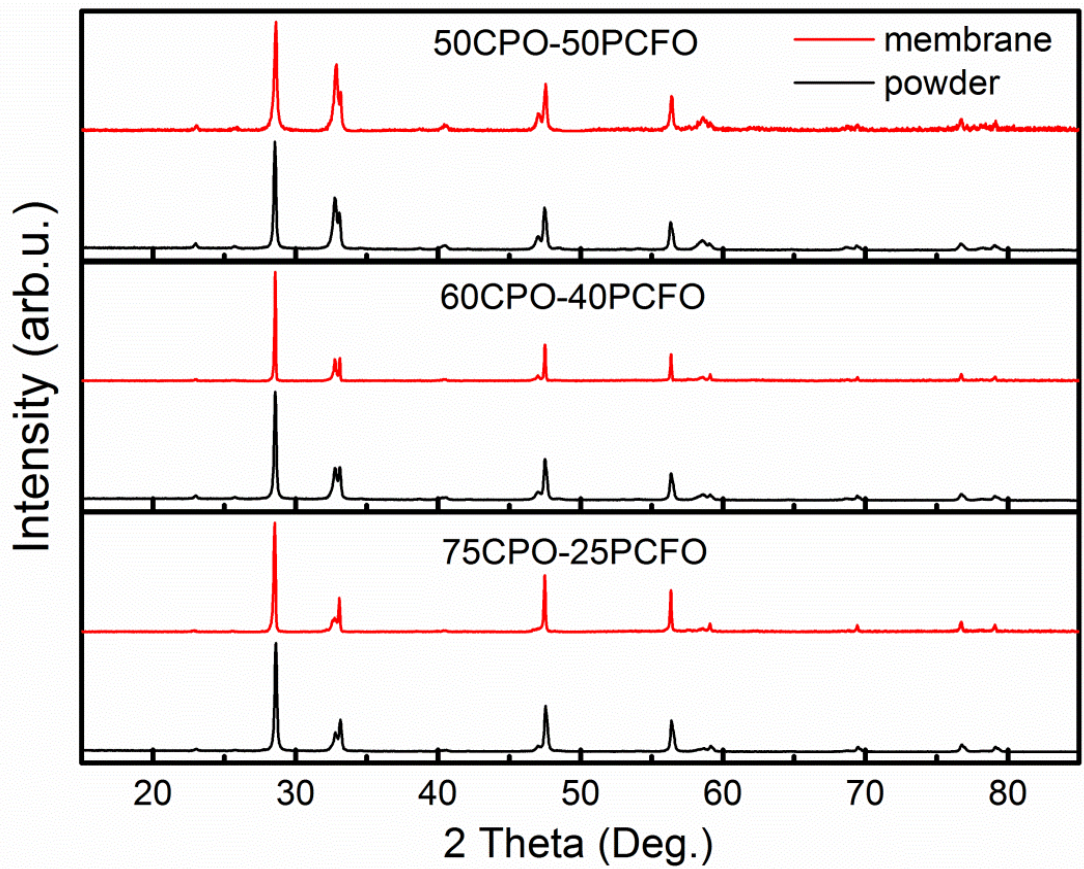

**Figure S3.** XRD patterns of (100-$x$)CPO-$x$PCFO ($x$ = 25, 40 and 50) powders after heated at 950 °C for 10 h and membranes after sintering at 1400 °C for 5 h.

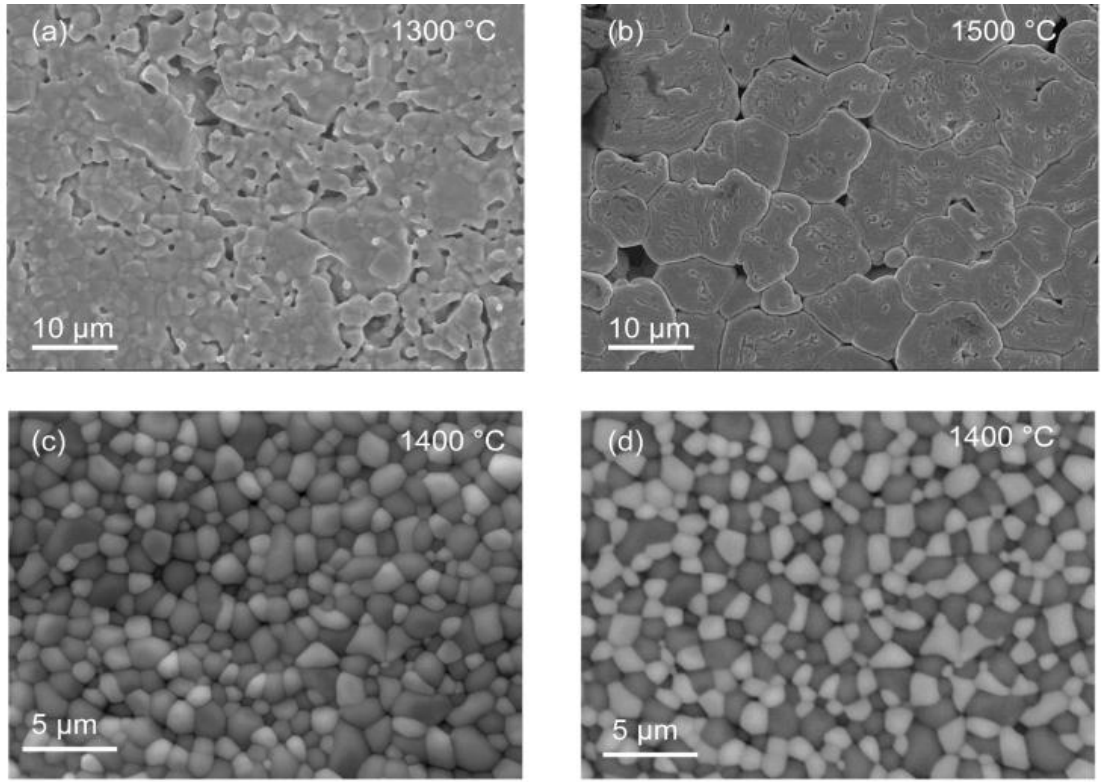

**Figure S4.** SEM images of 60CPO-40PCFO membranes sintered at (a) 1300 °C, (b) 1500 °C and (c) 1400 °C for 5 h; (d) BSEM images of 60CPO-40PCFO membrane sintered at 1400 °C for 5 h.

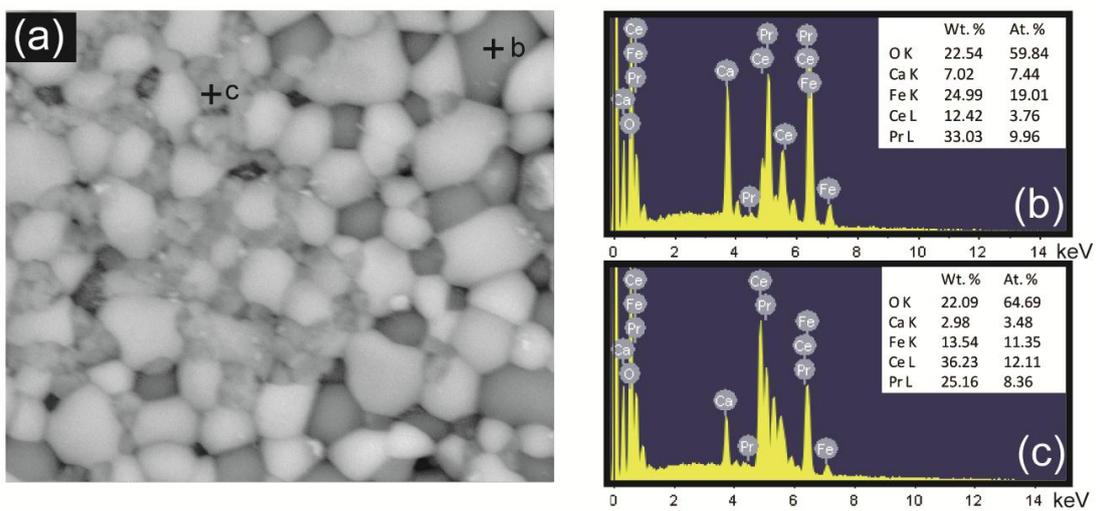

**Figure S5.** (a) EDXS image of 75CPO-25PCFO membrane after sintering at 1400 °C for 5 h; (b) EDXS of the black grain in 75CPO-25PCFO membrane; (c) EDXS of the grey grain in 75CPO-25PCFO membrane.

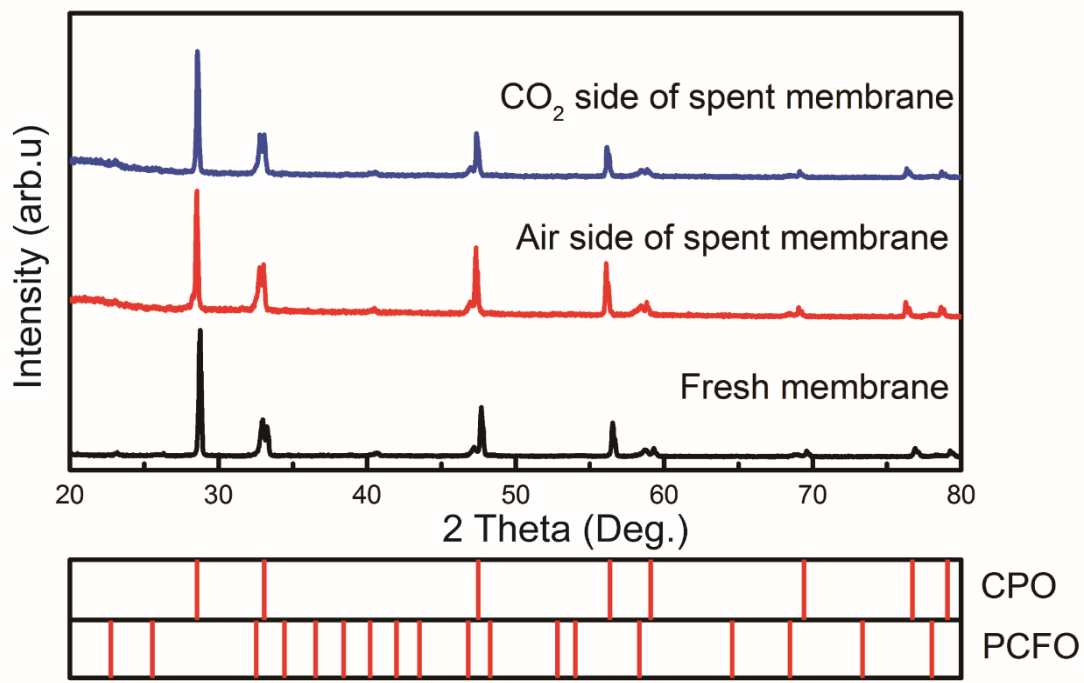

**Figure S6.** The XRD patterns of fresh and spent 60CPO-40PCFO dual-phase membranes after 100 h test under $CO_2$ atmosphere.

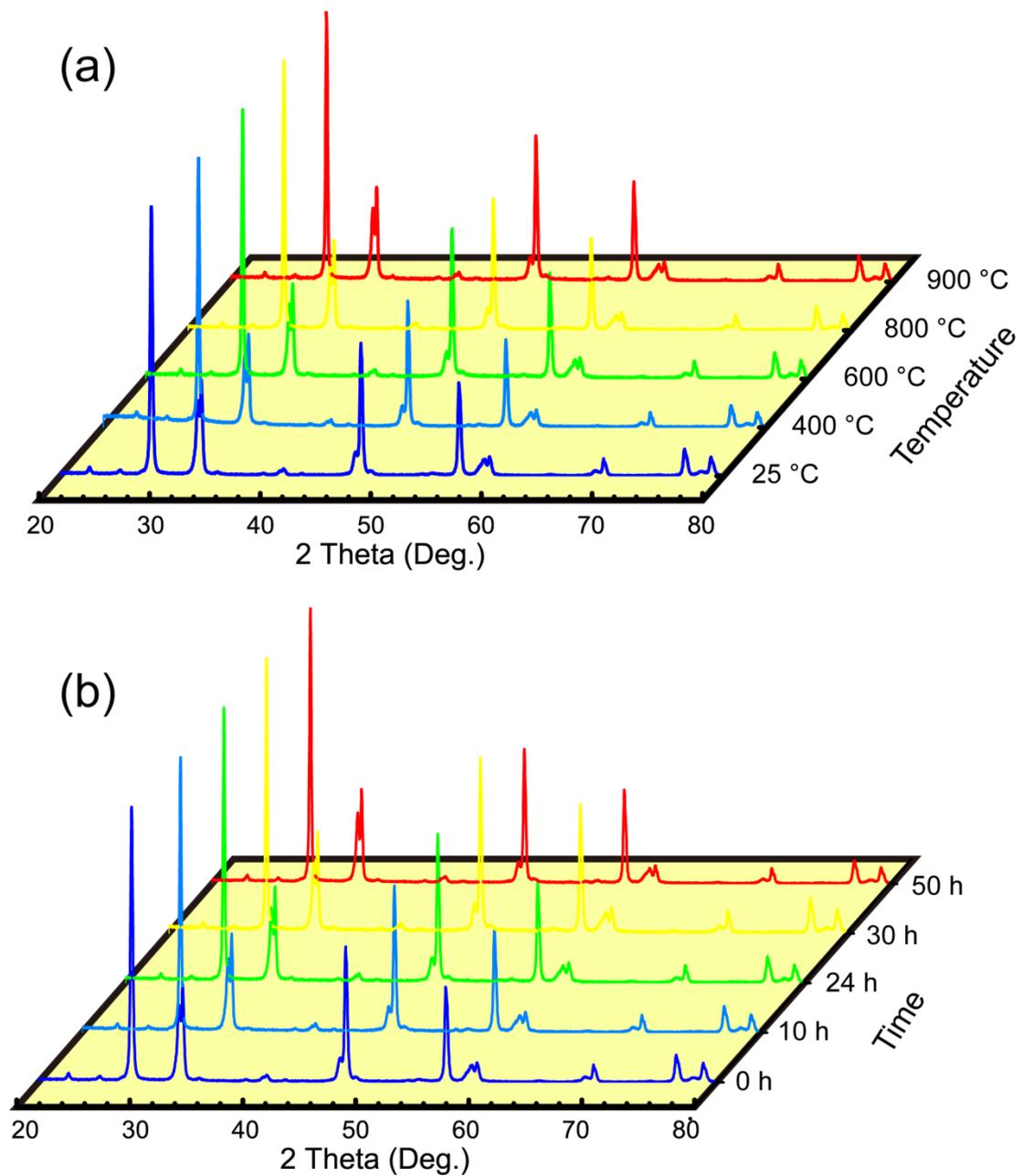

**Figure S7.** XRD patterns of 60CPO-40PCFO dual-phase membrane powder after exposed to pure Ar (a) at different temperatures and (b) at 950 °C for different hours.